\documentclass[trackchanges]{aastex701}

\usepackage{amsmath}
\usepackage{booktabs}

\begin{document}

\title{Physics of Circular Polarized Ion-Scale Waves in Hybrid Simulations of Alfvénic Fluctuations}

\author[orcid=0009-0007-9575-3900, gname='Hai Yang Harry', sname='Qian']{Hai Yang Harry Qian}
\affiliation{Physics Department, University of California, Berkeley, CA 94720-7300, USA}
\affiliation{Space Science Laboratory, University of California, Berkeley, CA 94720-7450, USA}
\email[show]{harry.qian26@berkeley.edu}  

\author[orcid=0000-0002-4625-3332,gname='Trevor A.', sname='Bowen']{Trevor A. Bowen}
\affiliation{Space Science Laboratory, University of California, Berkeley, CA 94720-7450, USA} 
\email{tbowen@berkeley.edu}

\author[orcid=0000-0001-7063-2511, gname='Carlos A.',sname='Gonzalez']{Carlos A. Gonzalez}
\affiliation{Department of Physics, The University of Texas at Austin, Austin, TX, USA}
\email{carlos.gonzalez1@austin.utexas.edu}

\author[0000-0002-1128-9685,sname=Sioulas ,gname=Nikos]{Nikos Sioulas}
\affiliation{Space Science Laboratory, University of California, Berkeley, CA 94720-7450, USA}
\email{nsioulas@g.ucla.edu}

\author[0000-0001-9202-1340,sname=Alfred ,gname=Mallet]{Alfred Mallet}
\affiliation{Space Science Laboratory, University of California, Berkeley, CA 94720-7450, USA}
\email{alfred.mallet@berkeley.edu}

\author[0000-0001-6038-1923,sname=Kristopher G. ,gname=Klein]{Kristopher G. Klein}
\affiliation{Department of Planetary Sciences and Lunar and Planetary Laboratory, University of Arizona, Tucson, Arizona 85721, USA}
\email{kgklein@arizona.edu}

\author[0000-0002-0497-1096,sname=Verscharen, gname=Daniel]{Daniel Verscharen}
\affiliation{Mullard Space Science Laboratory, University College London, Holmbury St Mary, Dorking, RH5 6NT, UK}
\email{d.verscharen@ucl.ac.uk}

\author[0000-0002-1989-3596]{Stuart D. Bale}
\affil{Physics Department, University of California, Berkeley, CA 94720-7300, USA}
\affil{Space Sciences Laboratory, University of California, Berkeley, CA 94720-7450, USA}
\affil{The Blackett Laboratory, Imperial College London, London, SW7 2AZ, UK}
\email{bale@berkeley.edu <mailto:email\%7Bbale@berkeley.edu>}


\begin{abstract}
Ion cyclotron waves (ICW) and fast magnetosonic/whistler waves (FMW) are fundamental electromagnetic modes at ion kinetic scales, yet their generation mechanisms and roles in plasma evolution remain poorly understood. We analyze a 2.5D hybrid simulation of broadband Alfv\'{e}nic fluctuations, where the proton velocity distribution is modeled as a sum of two bi-Maxwellian components: a thermal core and a drifting beam. Using wavelet-based wave identification, bi-Maxwellian VDF fitting, and the PLUME linear dispersion solver, we find that ICW behave as linear modes. Growth is intermittent, occurring when core temperature anisotropy builds up, and is driven mainly by the core (the beam contributes negligibly). Poynting flux analysis shows that ICW are predominantly forward-propagating, with a net energy flux ratio of $+1$ across all frequencies, consistent with the initial condition. FMW present a stark contrast: PLUME solutions often yield very small (near-zero) linear growth/damping rates. The species decomposition breaks down when $|\gamma/\omega_r| \gtrsim 0.368$, indicating that linear theory predicts these waves to be strongly damped and not describable by linear eigenmodes. Nevertheless, FMW are clearly observed in the wavelet helicity spectrogram, indicating that they are generated by nonlinear processes (e.g., parametric decay or phase steepening) and persist despite linear damping. The net energy flux ratio for FMW is close to $+1$ at low frequencies but decreases at higher frequencies, yet never reaches zero (net energy flow remains forward). These results demonstrate that ICW are linear, core-driven waves that transfer energy to the plasma, while FMW are heavily damped, nonlinearly generated waves.

\end{abstract}



\section{Introduction}

In the collisionless environment of the solar corona and solar wind, wave-particle interactions dominate over collisional processes, producing highly non-thermal distributions that deviate from Maxwellian equilibrium \citep{1982JGR....87...52M, 2012SSRv..172...23M, 2022PhRvL.129p5101B}. Observations show that solar wind temperature decreases with heliocentric distance more slowly than adiabatic expansion predicts, implying an active heating mechanism operating throughout the inner heliosphere \citep{https://doi.org/10.1029/94GL03273, 2013LRSP...10....2B}. Understanding these interactions is fundamental to explaining coronal heating and solar wind acceleration.

Ion cyclotron waves (ICW) are left-handed circularly polarized waves that resonate with ions through the normal cyclotron resonance condition $\omega - k_\parallel v_\parallel = \Omega_p$ where $\omega$ is the wave angular frequency, $k_\parallel$ is the wavenumber parallel to the background magnetic field, $v_\parallel$ is the ion velocity parallel to the magnetic field, and $\Omega_p$ is the proton gyrofrequency\citep{2013ApJ...773..163V}. They have been extensively studied as a dissipation mechanism that can explain preferential perpendicular heating and the acceleration of both protons and minor ions such as $O^{5+}$ (e.g., \cite{2009ApJ...696..591I, 2014ApJS..213...16C}). Fast magnetosonic/whistler waves (FMW), by contrast, are right‑handed circularly polarized when propagating parallel to the background field and can be driven by ion beams \citep{2026MNRAS.545f2089K, 2013ApJ...773..163V, 2024ApJ...961..142M}. Observational studies have identified both wave families in spacecraft data. Right‑handed FMW are often found near large‑scale structures such as the heliospheric current sheet and are highly intermittent, while left‑handed ICW are more commonly sampled and become increasingly frequent at closer heliocentric distances \citep{2022ApJ...924..112V, 2024ApJ...961..142M, 2026ApJS..284....4N}.


While the dissipative role of ICW is well established \citep{2014ApJS..213...16C,2024ApJ...972L...8B}, generation mechanisms for and impact of the FMW on solar wind plasma remain actively debated \citep{2021ApJ...914L..36G, 2024ApJ...961..142M}. One possibility for their relationship is that large-scale Alfv\'{e}nic fluctuations (observed in the solar wind) undergo nonlinear evolution, transferring energy to kinetic scales and generating proton beams. Indeed, spacecraft observations have revealed that proton beams are a persistent feature in the solar wind, particularly in collisionally young, fast wind streams \citep{2018ApJ...864..112A}. Hybrid simulations have demonstrated that both parametric instabilities \citep{araneda2008proton, matteini2010kinetics, 2023JPlPh..89b9008G} and phase steepening \citep{machida1987simulation, 2021ApJ...914L..36G} of Alfv\'{e}nic fluctuations produce field-aligned proton beams traveling at the Alfv\'{e}n speeds. These beams, in turn, can drive FMW, which then scatter and heat the plasma, potentially creating conditions favorable for ICW growth.

The cyclotron resonance condition enables efficient energy transfer between waves and particles, with counter-propagating waves proving particularly effective by allowing ions to diffuse in multiple directions in phase space \citep{2013PhRvL.110i1102K}. Theoretical work \citep{1999JGR...104.6759G} has established thresholds for these instabilities. For ICW, resonance with protons requires particles with parallel velocity opposite the wave phase velocity, and waves can be driven by either sufficient drift or temperature anisotropy \citep{2013ApJ...773..163V}. For FMW, on the other hand, resonance involves particles with positive parallel velocity, and the wave must avoid strong proton damping \citep{2013ApJ...773..163V, 2013ApJ...764...88V}. Recent observations have shown that ion-scale spectral steepening is associated with circular polarization \citep{2018ApJ...856...49W, 2019ApJ...884L..53W, 2024ApJ...972L...8B}, and that quasilinear heating rates can account for a significant fraction of the energy flux in the turbulent cascade rate \citep{2024ApJ...973...20S, 2024ApJ...972L...8B}.



Simulations provide a means to advance an understanding of wave generation and their role in dissipation \citep{2025ApJ...984..174O}. In this work, we analyze a high-cross-helicity 2.5D hybrid simulation of a parallel-propagating broadband Alfvénic fluctuation \citep{2024ApJ...963..148G}. We employ wavelet‑based identification of coherent waves, a technique commonly implemented in in situ spacecraft studies, and additionally fit the proton velocity distributions to a sum of two bi‑Maxwellian components (core and beam) to characterize the evolving core and beam populations. We identify both ICW and FMW in the simulation.
We use the Plasma in a Linear Uniform Magnetized Environment (PLUME) dispersion solver \citep{2025RNAAS...9..102K} to compute linear growth rates and identify which populations drive or damp each wave mode. Our results show that ICW behaves as linear modes that persist throughout the simulation, with local, intermittent growth primarily driven by the strong temperature anisotropy of the core proton population at the steepened fronts. For FMW, PLUME indicates no sustained linear growth; instead, the waves are predominantly damped. Notably, PLUME resolves finite growth/damping rates even at frequencies where the wavelet analysis does not identify coherent FMW signals. This suggests that the observed FMW are heavily damped linear modes and are likely nonlinearly generated.

The paper is organized as follows. Section \ref{model} describes the hybrid simulation model and initial conditions. Section \ref{methods} details our analysis methods, wavelet-based wave identification, fitting of proton velocity distributions, and linear dispersion analysis. Section \ref{analysis} presents our results, beginning with the temporal evolution of plasma parameters, followed by wavelet-based identification of ICW and FMW, and PLUME analysis of their growth, behavior, and species contributions. Finally, we examine the scale-dependent behavior of wave growth and damping. Section \ref{conclusion} summarizes our conclusions.

\section{Model and Simulation} \label{model}

The data used in this study are taken from a simulation performed with the CAMELIA hybrid particle-in-cell (PIC) code \citep{2018ApJ...853...26F}. The simulation treats protons as kinetic particles governed by the Vlasov–Maxwell equations, while electrons are modeled as a massless, isothermal fluid that maintains charge neutrality. This hybrid approach captures the essential ion kinetic physics (cyclotron resonance, temperature anisotropy, beam‑driven instabilities) without resolving electron kinetic scales, making it computationally tractable for the long durations and large domains required for turbulence studies.

We adopt standard hybrid simulation normalizations: lengths are normalized to the proton inertial length $d_i = c/\omega_p=v_A/\Omega_p$, where $\omega_p = (4\pi n e^2/m_i)^{1/2}$ is the proton plasma frequency; time is normalized to the inverse proton gyrofrequency $\Omega_p^{-1} = (eB_0/m_i c)^{-1}$; velocities are normalized to the Alfvén speed $v_A = B_0/\sqrt{4\pi n m_i}$; magnetic field is normalized to the background field $B_0$; densities are normalized to the background density $n_0$; the proton and electron $\beta$ are set equal, with $\beta = 8\pi n k_B T / B_0^2$.

The simulation domain is a square box of size \(L = 128\,d_i\), discretized into a \(1024 \times 1024\) grid points. The background magnetic field $B_0$ is oriented along the $x$-axis (in the simulation plane). The temporal data of the simulation fields analyzed in this paper are derived from the same broad parametric scan presented in \citet{2024ApJ...963..148G}. Initially, the plasma is homogeneous and isotropic, with protons loaded as a Maxwellian distribution (8000 particles per cell) with the plasma beta for both protons and electrons $\beta_{p,e}=0.5$. A broadband spectrum of outward‑propagating, left‑handed circularly polarized Alfvénic fluctuations is imposed \citep{2021ApJ...914L..36G, 2024ApJ...963..148G}. The initial pump wave is a purely one-dimensional fluctuation characterized by a parallel wavenumber spectrum $E_B(k_\parallel) \propto k_\parallel^{-2}$ over the range $k_\parallel d_i\in[0.049, 0.490]$, with a dominant mode $n=4$ ($k_0d_i\approx 0.196$).
The wave frequency \(\omega_0\) satisfies the cold‑plasma dispersion relation \(k_0^2 = \omega_0^2/(1-\omega_0)\), and the velocity perturbation follows the Walén relation \(\delta\mathbf{u} = -(\omega_0/k_0)\delta\mathbf{b}\). The magnetic field of the wave is given by  $\delta b_y = - \delta b_0 \sin{(\phi(k_0,x))}$ and $\delta b_z = \delta b_0 \cos{(\phi(k_0,x))}$, with $\delta b_0$ the amplitude of the wave normalized to the mean magnetic field magnitude $B_0$. The phase $\phi(k_0,x) = k_0 x + \epsilon \sum_{m=n_i, {m \neq n_0}}^{n_f}  \frac{k_0}{k_m} \cos{(k_m x + \phi_m)} $, where $\phi_m$ is a random phase between $ [0,2 \pi)$. The random $\phi_m$ ensures a broad, turbulent-like spectrum rather than a coherent monochromatic wave \citep{2021ApJ...914L..36G}.

\begin{figure}
    \centering
    \includegraphics[width=1\linewidth]{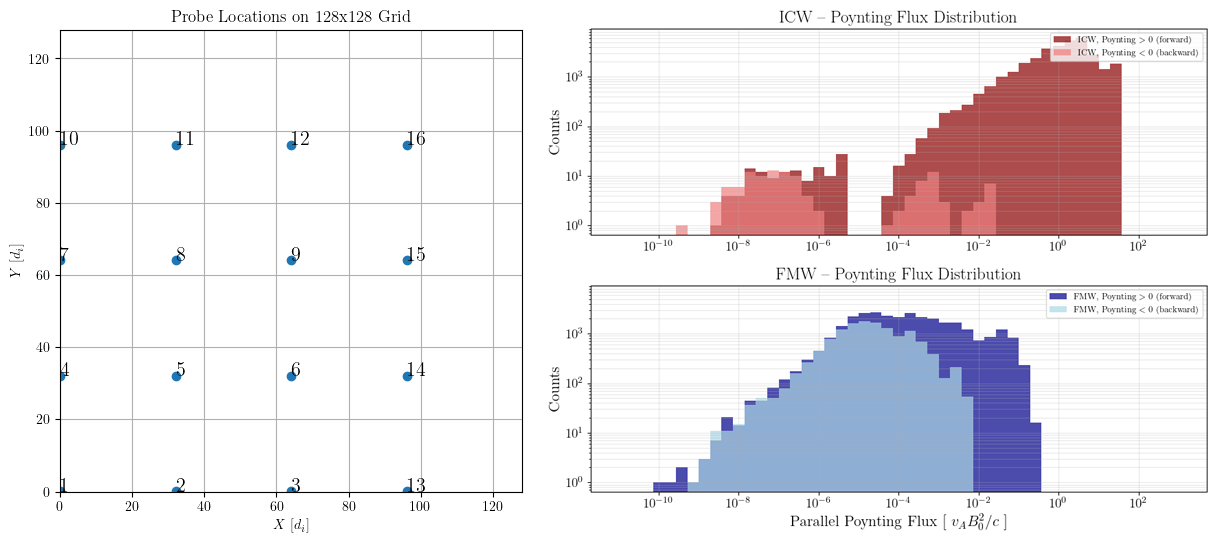}
    \caption{The left panel shows the probe locations on the simulation grid. The simulation domain is $128 \times 128$ cells in the grid plane, with spatial coordinates normalized to the proton inertial length $d_i$. These probes are distributed throughout the domain to capture spatial variability in plasma parameters and wave activity. The two right panels show the histogram of the parallel Poynting flux for ICW (red) and FMW (blue). Positive (negative) values indicate forward (backward) propagation.}
    \label{fig: probe location}
\end{figure}



To collect time‑series data, 16 fixed spatial probes are placed throughout the domain (Figure~\ref{fig: probe location} left). At each probe location we record the magnetic field $\mathbf{B}$, the electric field $\mathbf{E}$, the bulk velocity $\mathbf{u}$, and the particle density $n$ at a cadence of $\Delta t = 0.1\,\Omega_p^{-1}$ for a total duration of $300\,\Omega_p^{-1}$. To characterize the proton populations, we construct velocity distribution functions (VDFs) from the particle data collected at the same probe positions for every time step. For each particle, we first subtract the local bulk flow velocity to transform into the plasma frame: $\mathbf{v}' = \mathbf{v} - \mathbf{u}$. Here, “local” means the values measured at the probe location: $\mathbf{u}$ is the instantaneous mean particle velocity computed from all particles collected at that probe at that time, and the magnetic field $\mathbf{B}$ (used below) is the field recorded at the probe. The parallel velocity is obtained by projecting $\mathbf{v}'$ onto the local magnetic field direction: $v_\parallel = \mathbf{v}' \cdot \hat{\mathbf{B}}$. The perpendicular speed is defined as the magnitude of the velocity component perpendicular to the magnetic field: $v_\perp = \sqrt{|\mathbf{v}'|^2 - v_\parallel^2}$, which is always positive. We bin the particles in $(v_\parallel, v_\perp)$ space using $256 \times 256$ bins spanning $v_\parallel \in [-7.07\,v_A, 7.07\,v_A]$ and $v_\perp \in [0, 7.07\,v_A]$ to produce 2D histograms representing the phase‑space density.

\section{Methods} \label{methods}

\subsection{Wavelet-based identification of coherent waves}

Before performing wavelet analysis, the measured fields are transformed into the plasma rest frame. For each probe, we first compute the time‑averaged bulk velocity $\langle\mathbf{u}(\mathbf{r})\rangle$ over the entire simulation interval at that probe location (i.e., using the particle velocity time series from that probe). The electric field in the plasma frame is then obtained by removing the convective term associated with the bulk flow:
\begin{equation}
\mathbf{E'}=\mathbf{E}+\frac{\langle\mathbf{u}\rangle}{c}\times\mathbf{B}
    \label{Eq: E rest frame}
\end{equation}
using the same constant $\langle\mathbf{u}(\mathbf{r})\rangle$ for all time steps at that probe. This transformation isolates the wave properties from Doppler shifts caused by bulk motion. The magnetic field $\mathbf{B}$ is already measured in the probe frame, which coincides with the plasma frame (to first order in the non-relativistic limit, which is negligible for our simulation).

With the transformed fields $\mathbf{E'}$ and $\mathbf{B}$, we perform a wavelet transform to obtain the time-frequency spectrogram of the magnetic and electric field components. This transformation is normalized to units of energy and is defined as a convolution of the field data with a set of scaled wavelets, $\psi(s,\tau)$,
\begin{equation}
    W(s,t)=\sum_{i=0}^{N-1}\psi(\frac{t_i-\tau}{s})B(t_i).
    \label{Eq: convolution}
\end{equation}
The mother wavelet is given a fixed scale, $\psi(\tau)=\pi^{-1/4}e^{-i\omega_0\tau}e^{-\frac{\tau^2}{2}}$, where $\omega_0$ and $\tau$ are non-dimensional frequency and time parameters, respectively. We average the field over the wavelet's Gaussian envelope.  \citep{1992AnRFM..24..395F, 1998BAMS...79...61T, 2013SSRv..178..665D}. For this analysis, $\omega_0$ is chosen to be 6 \citep{2020ApJS..246...66B}. To avoid random fluctuations in the data, the local mean field is used instead. To calculate the mean field locally at a given scale, we average the field over the wavelet's Gaussian envelope. 
\begin{equation}
    B_{0j}(s,t)=\sum_{i=0}^{N-1}|\psi(\frac{t_i-\tau}{s})|B_j(t_i)
    \label{Eq: Gaussian envelope of the wavelet}
\end{equation}
where $j$ refers to the field component and $|\psi|=A_s\pi^{-\frac{1}{4}}e^{-\frac{\tau^2}{2}}$, where $A_s$ normalizes to unit energy \citep{2008PhRvL.101q5005H,2009ApJ...698..986P}.

The wavelet coefficients need to be transformed into a magnetic-field-aligned coordinate system, defined as ($\hat{B_{\perp 1}}$, $\hat{B_{\perp 2}}$, $\hat{B_0}$). $\hat{B_0}$ is the mean magnetic field, where $\hat{B_{\perp 1}}$ is calculated as the cross product of the vector in the direction of maximum variance with $\hat{B_0}$, and $\hat{B_{\perp 2}}$ ensures a closure for a right-handed coordinate system, namely $\hat{B_{\perp 1}} \times \hat{B_{\perp 2}} = \hat{B_0}$. We then define parameters analogous to Stokes parameters that describe the polarization state of electromagnetic radiation:
\begin{equation}
    S_0(f,t)=B_{\perp 1}^2 + B_{\perp 2}^2
    \label{Eq: Stokes 0}
\end{equation}
\begin{equation}
    S_1(f,t)=B_{\perp 1}^2 - B_{\perp 2}^2
    \label{Eq: Stokes 1}
\end{equation}
\begin{equation}
    S_2(f,t)=2\Re(B_{\perp 1}B_{\perp 2}^*)
    \label{Eq: Stokes 2}
\end{equation}
\begin{equation}
    S_3(f,t)=-2\Im(B_{\perp 1}B_{\perp 2}^*)
    \label{Eq: Stokes 3}
\end{equation}
Equation \ref{Eq: Stokes 3} expresses the circular polarization of the perpendicular magnetic field along the parallel direction \citep{1994STIA...9560772M, 2009AnGeo..27.3967N, 2010ApJ...709L..49H}. In particular, the normalized circular polarization (or reduced helicity):
\begin{equation}
    \sigma(f,t)=\frac{S_3}{S_0}=\frac{-2\Im(B_{\perp 1}B_{\perp 2}^*)}{B_{\perp 1}^2 + B_{\perp 2}^2}
    \label{Eq: polarization}
\end{equation}
which equals $+1$ for left-handed polarization (ICW) and $-1$ for right-handed polarization (FMW). To reduce the effect of turbulence with instantaneous polarization, the quantities are time-averaged using a two-e-folding Gaussian envelope of the Morlet wavelet.

For the transformed electric fields $\mathbf{E'}$, we perform a wavelet transform on their components, identical to the procedure used for the magnetic field. The resulting electric wavelet coefficients are then projected into the same magnetic-field-aligned coordinate system ($\hat{B_{\perp 1}}$, $\hat{B_{\perp 2}}$, $\hat{B_0}$) that was constructed for the magnetic field. This ensures that the perpendicular components of both fields are expressed in a common basis. To classify a parallel propagating wave, a selection criterion is used to separate the background turbulence from the wave events. For each scale, a wave event is identified when the $|\sigma_{B}| \text{ and }|\sigma_{E}| >0.7$ and extends to adjacent times while $|\sigma_{B}| \text{ and }|\sigma_{E}|>0.5$ \citep{2020ApJS..246...66B}. To verify if the waves are ICW or FMW, their phase speed is calculated and used to estimate the associated wavevector $k_\parallel$,
\begin{equation}
    \frac{E_{\perp1}}{B_{\perp2}}=-\frac{E_{\perp2}}{B_{\perp1}}=\frac{v_{\text{phase}}}{c}=\frac{\omega}{k_\parallel c}
    \label{Eq: phase w/k}
\end{equation}
where $E_{\perp1}$, $E_{\perp2}$, $B_{\perp1}$, and $B_{\perp2}$ are the wavelet coefficients during the coherent times in the $\perp_1$ and $\perp_2$ directions defined in the magnetic-field-aligned coordinate system. $\omega$ is the frequency at which the wavelet-transformation resolved the data. To assess proper fit, the coefficient of determination ($R^2$) is computed to determine whether the data points fit a linear model. Only data with $R^2>0.85$ are considered in order to select only parallel propagating waves. In addition to this requirement, the duration of the wave also needs to satisfy the condition that it must be at least one-$e$-folding length at a given frequency.

\begin{figure}
    \centering
    \includegraphics[width=0.5\linewidth]{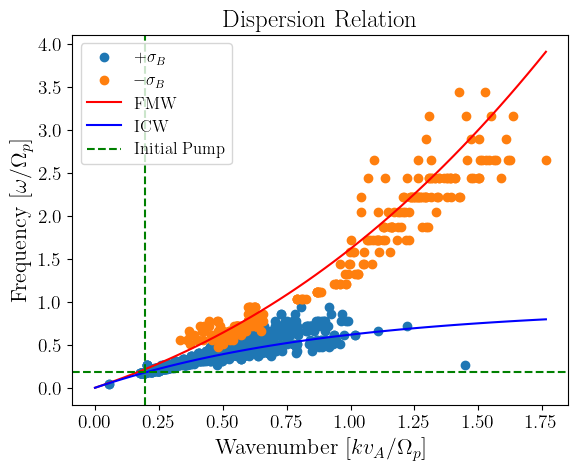}
    \caption{Theoretical cold plasma dispersion relation for ion-scale waves in a magnetized plasma with parameters corresponding to the simulation conditions. The frequency $\omega$ (normalized to proton gyrofrequency $\Omega_p$) is plotted against wavenumber $k$ (normalized to inverse proton inertial length $v_A/\Omega_p$). The linear dispersion consists of two distinct branches: the left‑handed (ICW) branch lies slightly below the right‑handed (FMW) branch at all wavenumbers; both increase monotonically with $k$ and approach $\omega\approx\Omega_p$ as $kv_A/\Omega_p\to1$. $+\sigma$ refers to the left-handed ICW, and the $-\sigma$ refers to the right-handed FMW. Right-handed fluctuations align with the FMW branch; left-handed fluctuations align with the ICW branch, confirming mode identification. The dashed green lines represent the initial pump's central parameters.
}
    \label{fig: dispersion}
\end{figure}

After computing $v_{\text{phase}}$ for each coherent wave event, we obtain the corresponding wavenumber as $k=\omega/v_{\text{phase}}$. The resulting dispersion points $(\omega,k)$ are then plotted and compared with the theoretical cold plasma dispersion curves for parallel-propagating ICW and FMW (Figure~\ref{fig: dispersion}). Good agreement between the simulation data and the theoretical branches confirms that the observed waves correspond to the expected linear eigenmodes. This step validates wave identification using the reduced magnetic helicity criterion.


To quantify the direction of wave energy transport, we compute the parallel component of the Poynting vector from the wavelet coefficients of the electric and magnetic fields: \(S_{\parallel} = \frac{1}{2}\Re\{(\mathbf{E} \times \mathbf{B}^*)_{\parallel}\}\). Figure~\ref{fig: probe location} (right) shows the distribution of the magnitude of Poynting flux for each wave. 

From the reduced magnetic helicity \(\sigma_B\), we identify coherent wave packets as ICW (\(\sigma_B > 0.7\)) or FMW (\(\sigma_B < -0.7\)). For each frequency scale and each wave type, we sum the positive (forward) and negative (backward) values of \(S_{\parallel}\) to obtain the total forward energy \(E_f\) and total backward energy \(E_b\). The net energy flux ratio \(\mathcal{R} = (E_f - E_b)/(E_f + E_b)\) then varies between \(+1\) (purely forward) and \(-1\) (purely backward). For ICW, \(\mathcal{R}\) is consistently \(+1\) across all resolved frequencies, confirming that they are entirely forward‑propagating, consistent with the initial outward pump. For FMW, \(\mathcal{R}\) is close to \(+1\) at low frequencies (\(\omega/\Omega_p \lesssim 0.2\)) and gradually decreases at higher frequencies, yet it never reaches zero or a negative value (with a minimum value of $\mathcal{R}=0.1$ at $\omega/\Omega_p=2$). This indicates that while the energy flux balance shifts somewhat towards backward propagation at larger frequencies, the net energy flow remains forward at all scales. 

\subsection{Bimaxwellian VDF fitting}
To extract the quantitative parameters, we fit each VDF with a bi-Maxwellian model consisting of a proton core and a proton beam population,
\begin{equation}
    f(v_{\parallel}, v_{\perp})=\sum_{j=c,b}\frac{n_j}{\pi^{3/2}w_{\parallel, j}w_{\perp, j}^2}v_{\perp}e^{-\frac{(v_{\parallel}-V_j)^2}{w_{\parallel,j}^2}-\frac{v_{\perp}^2}{w_{\perp,j}^2}}
    \label{Eq: bixmax}
\end{equation}
where $n_j$ represents the species' density, $V_j$ represents the species' field-aligned drift speed, and $w_{\parallel/\perp,j}$ represents the species' thermal velocity (parallel and perpendicular direction with respect to the local mean magnetic field). The core proton population and the beam proton population are denoted by subscripts $c$ and $b$, respectively. This approach allows us to track the evolution of core and beam properties throughout the simulation.

We perform nonlinear least-squares fits of the model in Eq.~\ref{Eq: bixmax}. The fitting is performed using constrained nonlinear least-squares minimization with the \textbf{SLSQP} (Sequential Least Squares Programming) algorithm implemented in \textit{scipy.optimize.minimize}. The fit parameters are ${n_c, n_b, w_{\perp,c}, w_{\perp,b}, w_{\parallel,c}, w_{\parallel,b}, V_{c}, V_{b}}$. To ensure physically meaningful results, we apply bounds on each parameter and enforce the constraint $n_c > n_b$ (core density exceeds beam density) and the temperature anisotropy throughout the simulation cannot exceed 10 ($T_\perp/T_\parallel <10$). At each timestep, we initialize the fit using the converged parameters from the previous timestep and add a small random perturbation to avoid local minima. This sequential approach provides stability while allowing the fit to track temporal evolutions.

We assess the fitting quality through several metrics. The reduced chi-squared statistic $\chi^2_\nu = \sum (f_{\text{data}} - f_{\text{fit}})^2 / \nu$ (where $\nu$ is the degrees of freedom) quantifies the goodness of fit. For time steps where the fit failed to converge or produced nonphysical parameters, we flag and exclude those points from subsequent analysis (this is rare, with only $\lesssim 5\%$ of time steps failing the fit). We also estimate parameter uncertainty from the covariance matrix, computed as $\text{cov} = \chi^2_\nu, (\mathbf{J}^T \mathbf{J})^{-1}$, where $\mathbf{J}$ is the Jacobian matrix approximated at the solution. 

\subsection{PLUME dispersion characterization}
The fitted parameters are used as inputs for the Plasma in a Linear Uniform Magnetized Environment (PLUME) dispersion solver \citep{2025RNAAS...9..102K}. PLUME solves the linear Vlasov–Maxwell dispersion relation for a hot, magnetized plasma where each species is represented by a drifting bi‑Maxwellian distribution. The wavevector is defined in terms of components perpendicular and parallel to the background magnetic field, normalized to the reference gyroradius \(\rho_R = v_{\parallel R}/\Omega_R\). For the runs presented here, the perpendicular component is fixed at \(k_\perp \rho_R = 10^{-4}\) to enforce nearly parallel propagation, while the parallel component \(k_\parallel\) is scanned over the range of interest. Global reference parameters are shared across all timesteps: the reference parallel plasma beta \(\beta_{\parallel R}\) (taken from the core proton) and the reference parallel thermal speed \(v_{\parallel R}/c = 10^{-4}\). The reference species is the core proton. PLUME then computes the complex frequency \(\omega_r + i\gamma\) as a function of \(k_\parallel\). Wave polarization (left‑ or right‑handed) is determined from \(\Im\{E_y/E_x\}\). Following \citet{1998ApJ...500..978Q}, PLUME provides species‑resolved growth/damping rates \(\gamma_s\), valid for \(|\gamma/\omega_r| \ll 1\), such that \(\gamma_{\text{tot}} = \gamma_c + \gamma_b + \gamma_e\). This allows identification of which population drives or damps each mode. PLUME converges for all time steps, enabling a complete linear stability analysis across the simulation.

\section{Data Analysis} \label{analysis}
\subsection{Wavelet-based Analysis}

\begin{figure}
    \centering
    \includegraphics[width=1 \linewidth]{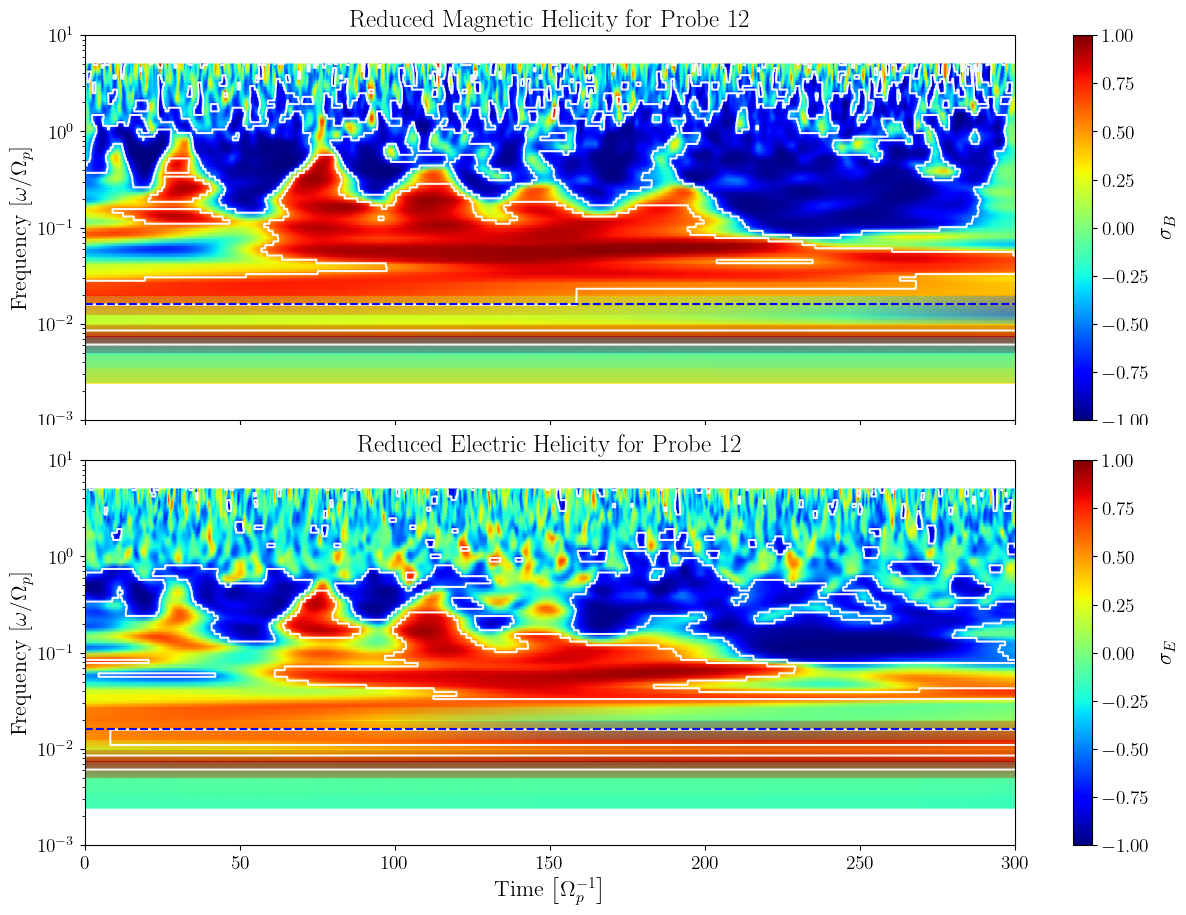}
    \caption{Reduced magnetic (top) and electric (bottom) helicity spectrograms. The reduced magnetic and electric helicity $\sigma_{B,E}$ identifies the polarization of the wave. Both quantities are normalized to $\pm 1$, with positive values (red) indicating left-handed polarization and negative values (blue) indicating right-handed polarization. The time-frequency structure reveals the presence of both wave populations. The dashed blue line in both panels indicates the central frequency of the initial broadband Alfvén wave pump. The white contour shows the identified waves.}
    \label{fig: sigmaBE}
\end{figure}

The application of wavelet-based polarization analysis methodologies \citep{2018ApJ...856...49W,2020ApJS..246...66B, 2024ApJ...973...20S} to our simulation data \citep{2024ApJ...963..148G} provides a way to understand the ion-scale wave dynamics. While reduced magnetic helicity can identify coherent wave packets in both spacecraft and simulation data, the presence of the solar wind flow in spacecraft measurements can Doppler-shift the intrinsic wave polarization, making it ambiguous whether an observed left-handed wave corresponds to an ICW or a Doppler-shifted FMW \citep{2020ApJ...899...74B}. In our simulation, we have full electric-field information in the plasma rest frame, enabling us to resolve this ambiguity and reliably classify waves as ICW or FMW. Figure~\ref{fig: sigmaBE} reveals the generation of wave packets, characterized by strong circular polarization ($|\sigma| \approx 0.85$), which occupy a well-defined frequency band in the plasma frame: $\omega/\Omega_p\in[0.03,0.52]$ for ICW and $\omega/\Omega_p\in[0.09,2]$ for FMW. This frequency range extends from the initial pump frequency to ion-kinetic scales, matching the range over which ion-scale waves are observed by spacecraft. The simulated waves thus serve as a valuable tool for interpreting in situ observations.

A central finding of this analysis is the clear presence of both coherent parallel-propagating ion-scale wave modes: the left-handed ICW and the right-handed FMW mode (Figure~\ref{fig: sigmaBE}). As illustrated in Figure~\ref{fig: dispersion}, the data points, colored by their reduced magnetic helicity, are plotted against the theoretical cold plasma dispersion curves for ICW and FMW waves. The good alignment of the positive $\sigma_B$ (left-handed) data with the ICW branch and the negative $\sigma_B$ (right-handed) data with the FMW branch verifies that the simulation generates both ion-scale electromagnetic modes propagating parallel to the mean magnetic field. This clear separation and agreement with linear theory validate the wavelet-based identification methodology and confirm that the simulation accurately captures the linear wave physics of a magnetized, collisionless plasma.
\subsection{Bimaxwellian VDF Analysis}

\begin{figure}
    \centering
    \includegraphics[width=0.9\linewidth]{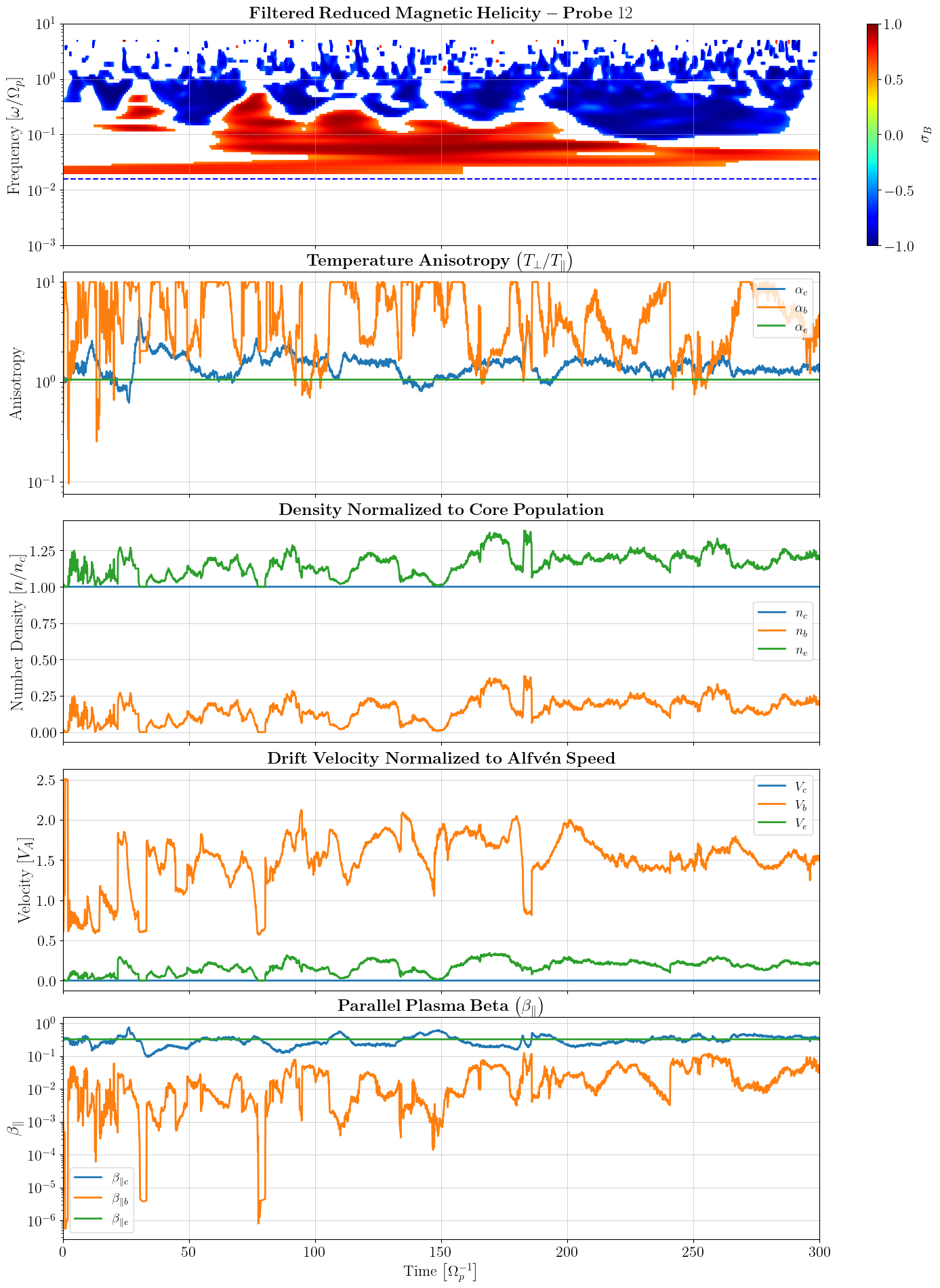}
    \caption{Characteristic plasma parameters during the entire simulation interval for probe 12. Key parameters include (from top to bottom panel): filtered reduced magnetic helicity ($\sigma_B$), temperature anisotropy for each species ($T_\perp/T_\parallel$), density for each species normalized to the core population ($n/n_c$), drift velocity normalized to Alfvén local speeds and with respect to the core population ($v_d/v_A$), and parallel plasma beta for each species. The dashed blue line on the filtered reduced magnetic helicity spectrogram shows the central frequency of the initial pump.}
    \label{fig: temporal series}
\end{figure}

Figure \ref{fig: temporal series} shows the extracted and computed plasma parameters over time for probe 12. Each panel, except the topmost one (filtered reduced magnetic helicity spectrogram), shows the plasma parameter for each species (core, beam, electron). The filtered reduced magnetic helicity spectrogram retains only data classified as wave events ($|\sigma_{B,E}|>0.7$) and extends to adjacent times with $|\sigma_{B,E}|>0.5$.

The beam temperature anisotropy, $\alpha_b=T_{\perp,b}/T_{\parallel,b}$ exhibits rapid, non-physical spikes throughout the simulation (Figure~\ref{fig: temporal series}, second panel). These spikes occur when the beam becomes either very low in density or extremely cold in the parallel direction, making the anisotropy ratio highly sensitive to small statistical fluctuations in the velocity distribution function and to numerical noise in the fitting routine. Under such conditions, the fitting algorithm returns anomalously small values of $\omega_{\parallel,b}$ leading to the observed spikes. These spikes do not represent real measurements but rather numerical artifacts of the bi‑Maxwellian fit when the beam population is poorly resolved.

The core density is set to 1, as densities are normalized to it. The electron density and beam density have the same shape. The electron density is determined by quasineutrality, so its trend follows the beam density trend (since the core density is constant). The beam and core are two components of the same underlying distribution. Their densities are anti-correlated: when $n_b$ increases, $n_c$ must decrease. However, $n_c$ is fixed by normalization. This normalization choice hides the anti-correlation; in physical units, an increase in beam density would necessarily decrease core density. Thus, the constant $n_c$ is an artifact of normalization, not a physical invariance.

\begin{figure}
    \centering
    \includegraphics[width=1\linewidth]{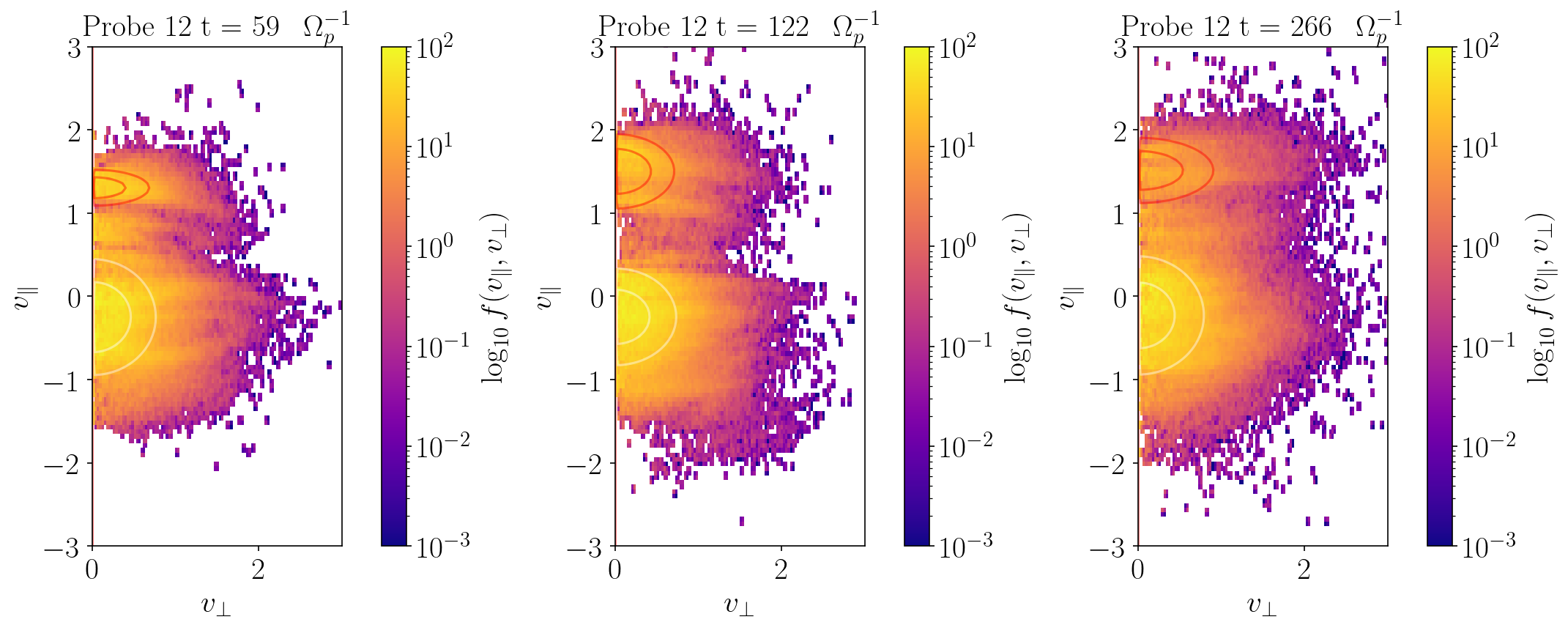}
    \caption{Proton velocity distribution functions (VDFs) at three selected times (top: $t = 59\Omega_p^{-1}$, middle: $t = 122 \Omega_p^{-1}$, bottom: $t = 266\Omega_p^{-1}$). The color images show the raw 2D histograms of particle counts in $(v_\parallel, v_\perp)$ space. Overlaid white contours represent the best‑fit bi‑Maxwellian model for the core population, and the red contours represent the best‑fit bi‑Maxwellian model for the beam population obtained from the non‑linear least‑squares fit. The fitted parameters (densities, thermal speeds, and drifts) are reported in Table~\ref{tab: fits}. Note that the fits are performed on the original simulation data without additional normalization (e.g., the core density is not forced to 1, nor is the core drift set to zero).}
    \label{fig: fits}
\end{figure}

The fact that the bi-Maxwellian fits reproduce the observed VDFs well, as shown by the contours in Figure~\ref{fig: fits} and summarized in the Table~\ref{tab: fits}) validates our fitting method and confirms that the core-plus-beam decomposition captures the essential features of the proton velocity distributions. Consequently, the extracted parameters, including the beam temperature anisotropy, are valid where the beam density is sufficiently high. The high beam temperature anisotropy can therefore be interpreted as genuine: it indicates that the beam population has undergone strong perpendicular heating relative to its parallel temperature. Their associated $\alpha_b$ at times $t=59\Omega_p^{-1}, 122\Omega_p^{-1}, 266\Omega_p^{-1}$ are $\alpha_b=10.4, 1.20, 6.63$ respectively.

\begin{table}
\centering
\caption{Fitted bi-Maxwellian parameters at selected times. The quantities are defined as follows: $n_c$ and $n_b$ are core and beam densities (normalized to core density); $w_{\perp c}$, $w_{\perp b}$ are perpendicular thermal speeds; $w_{\parallel c}$, $w_{\parallel b}$ are parallel thermal speeds; $V_c$ and $V_b$ are parallel drift speeds (normalized to $v_A$). All quantities are dimensionless in simulation units.}
\begin{tabular}{ccccccccc}
\toprule
Time $[\Omega_p^{-1}]$ & $n_c$ & $n_b$ & $w_{\perp c}$ & $w_{\perp b}$ & $w_{\parallel c}$ & $w_{\parallel b}$ & $V_c$ & $V_b$ \\
\midrule
59  & 1.00 $\pm$ 0 & 0.19 $\pm$ 0.01 & 0.72 $\pm$ 0.01 & 0.71 $\pm$ 0.01 & 0.66 $\pm$ 0.01 & 0.22 $\pm$ 0.01 & 0.00 $\pm$ 0 & 1.57 $\pm$ 0.01\\
122 & 1.00 $\pm$ 0 & 0.21 $\pm$ 0.02 & 0.73 $\pm$ 0.01 & 0.69 $\pm$ 0.01 & 0.58 $\pm$ 0.01 & 0.63 $\pm$ 0.01 & 0.00 $\pm$ 0 & 1.72 $\pm$ 0.01\\
266 & 1.00 $\pm$ 0 & 0.13 $\pm$ 0.01 & 0.78 $\pm$ 0.01 & 0.85 $\pm$ 0.01 & 0.73 $\pm$ 0.01 & 0.33 $\pm$ 0.01 & 0.00 $\pm$ 0 & 1.73 $\pm$ 0.01\\
\bottomrule
\end{tabular}
\label{tab: fits}
\end{table}

\subsection{PLUME Analysis}
Following the methodology outlined by \citet{2021ApJ...909....7K}, we perform a comprehensive linear stability analysis of the plasma conditions in our PLUME simulation. The VDFs obtained from the bi‑Maxwellian fitting procedure capture the essential kinetic features of the proton core and beam populations. These fitted parameters are used as inputs to PLUME to solve the hot‑plasma dispersion relation and compute the growth and damping rates (\(\pm\gamma/\Omega_p\)) of wave modes across a range of parallel wavenumbers \(k_\parallel\) (with the perpendicular component \(k_\perp = 0.001\, k \rho_p\) for nearly parallel propagation). PLUME scans a user‑defined rectangle in the complex frequency plane \(\overline{\omega} = \overline{\omega}_r + i\gamma\), searching for minima of \(|\det\overline{\Lambda}(k,\overline{\omega})|\) where \(\overline{\Lambda}\) is the wave matrix constructed from the dielectric tensor. The scan covers real frequencies in \(|\overline{\omega}_p| \in [0.002, 20]\) (i.e., both positive and negative sides) and imaginary parts in \(|\gamma|/\Omega_p \in [2\times10^{-6}, 1]\) (again considering both growth and damping). Once candidate roots are identified, a gradient‑descent solver refines them to high precision, yielding the complex frequency \(\omega+i\gamma\) for each mode. The solver retains physically admissible solutions, those with \(\gamma/\Omega_p>0\) (growing) or only weakly damped. Results are presented in time‑frequency spectrograms with the growth/damping rate shown by the color scale.

For the solutions shown, the growth and damping rates are capped at $|\gamma|/{\Omega_{p}}>10^{-3}$. According to linear theory, a wave mode undergoes exponential growth proportional to $\exp(\tau\gamma/\Omega_{p})$, where $\tau$ is a characteristic time. A value of $\tau\cdot\gamma/\Omega_p\gtrsim1$ signifies that the wave amplitude can increase by a factor of $e$ or more during the characteristic time, implying that linear growth is substantial. Conversely, if $\tau\cdot\gamma/\Omega_p\ll1$, it indicates that linear growth is insufficient to account for the significant wave amplification within the characteristic time. For our purposes, let the characteristic time constant be the simulation time. For a linear growth to be plausible, it requires that $\gamma/\Omega_{p}$ needs to be at least $\gamma/\Omega_{p}\gtrsim1/\tau$ or on the order of $10^{-3}$.

\begin{figure}
    \centering
    \includegraphics[width=1\linewidth]{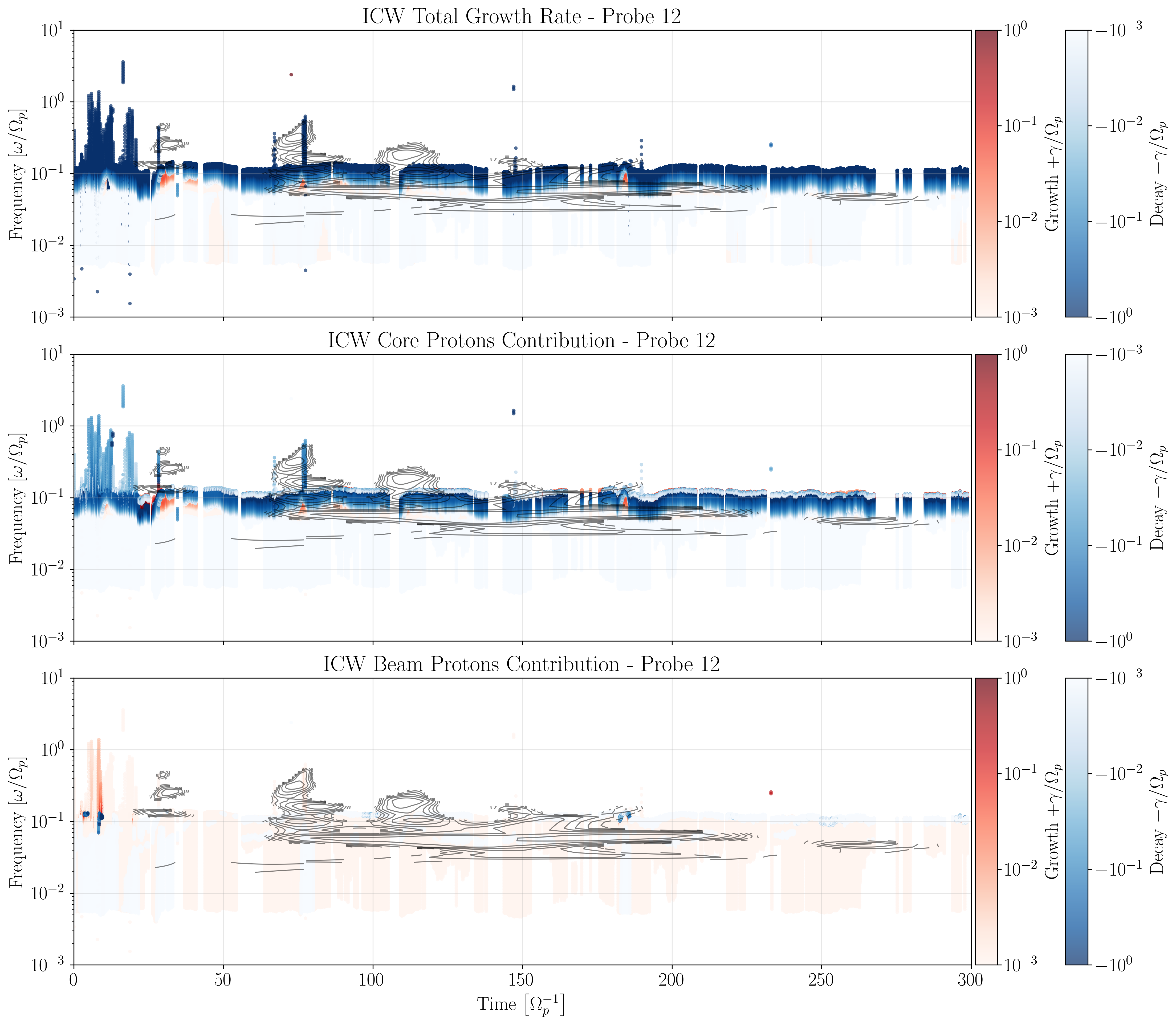}
    \caption{PLUME-derived ICW growth rates decomposed by species. The total growth rate $\gamma/\Omega_p$ (top) shows the total contribution by summing all species' contributions. The core contribution (second panel) is the major contribution toward the total for both growing and damping. The beam (bottom) plays a minor role, as its contribution is low. The contours represent the reduced magnetic helicity $\sigma_B$ from the simulation with values $\sigma_B>0.7$.}
    \label{fig: ICW gamma}
\end{figure} 

Figure~\ref{fig: ICW gamma} shows the growth and damping contribution from each species computed by PLUME for the ICW branch. From the total contribution ($\gamma_{\text{tot}}=\gamma_c+\gamma_b+\gamma_e$), positive growth occurs only during isolated intervals (e.g., at around $t=30, 77, 187$ $\Omega_{p}^{-1}$), while at most other times the wave is damped. This illustrates that ICW are intermittently generated, indicating that the plasma conditions required for ICW instability are met only occasionally. The black outline in Figure~\ref{fig: ICW gamma} is the contour of the wave events classified by the wavelet analysis. We observe that the wave event frequencies estimated by wavelet analysis and PLUME are relatively well aligned. 

Before ICW growth, we observe a systematic increase in $\alpha_c$ (Figure~\ref{fig: temporal series}, second panel), indicating that free energy accumulates in the core. This buildup of anisotropy may be associated with phase steepening of Alfvénic fluctuations, which generates localized regions of strong temperature anisotropy and field‑aligned beams \citep{2021ApJ...914L..36G, 2024ApJ...963..148G}. Intermittent ICW are then excited at these steepened fronts where the core anisotropy exceeds the instability threshold. During the growth interval itself, $\alpha_c$ decreases, consistent with the ICW extracting energy from the core anisotropy via cyclotron resonance. Following the growth event, $\alpha_c$ continues to relax, suggesting the plasma approaches a marginally stable state, as described in \citet{1996JGR...10111055I} and \citet{2014ApJS..213...16C}. These results establish ICW as linear modes driven intermittently by the core, consistent with cyclotron resonance theory \citep{2013ApJ...773..163V, 2014ApJS..213...16C}.

At early simulation times ($t<25\Omega_p^{-1}$), the PLUME output exhibits erratic growth/damping rates at high frequencies. These spurious values arise from numerical artifacts during the initial transient phase. The plasma is still evolving from the prescribed initial condition, and the bi‑Maxwellian fits are likely to be poorly constrained (due to the tenuous, emerging beam), introducing noise into the input parameters. High‑frequency modes are particularly sensitive to small fluctuations in density, drift, and temperature anisotropy, causing the solver’s automatic complex‑frequency scan to occasionally latch onto non‑physical roots. As the simulation proceeds, the turbulence develops, the fitted parameters stabilize, and the gamma values become physically consistent (e.g., intermittent ICW growth). Therefore, we exclude the first $\sim25\Omega_p^{-1}$ time from quantitative analysis of growth/damping rates, focusing on the later, reliable portion of the spectrogram.

To further understand which species is driving the growth, the individual contribution to the growth/damping is displayed (Figure~\ref{fig: ICW gamma}). The core makes a significant contribution to the overall growth compared to the beam. The beam contribution is negligible at all times (typically less than $5\%$). This confirms that the beam, despite carrying substantial free energy (evidenced by its high drift speed $V_b > v_A$), does not directly drive ICW.

\begin{figure}
    \centering
    \includegraphics[width=1\linewidth]{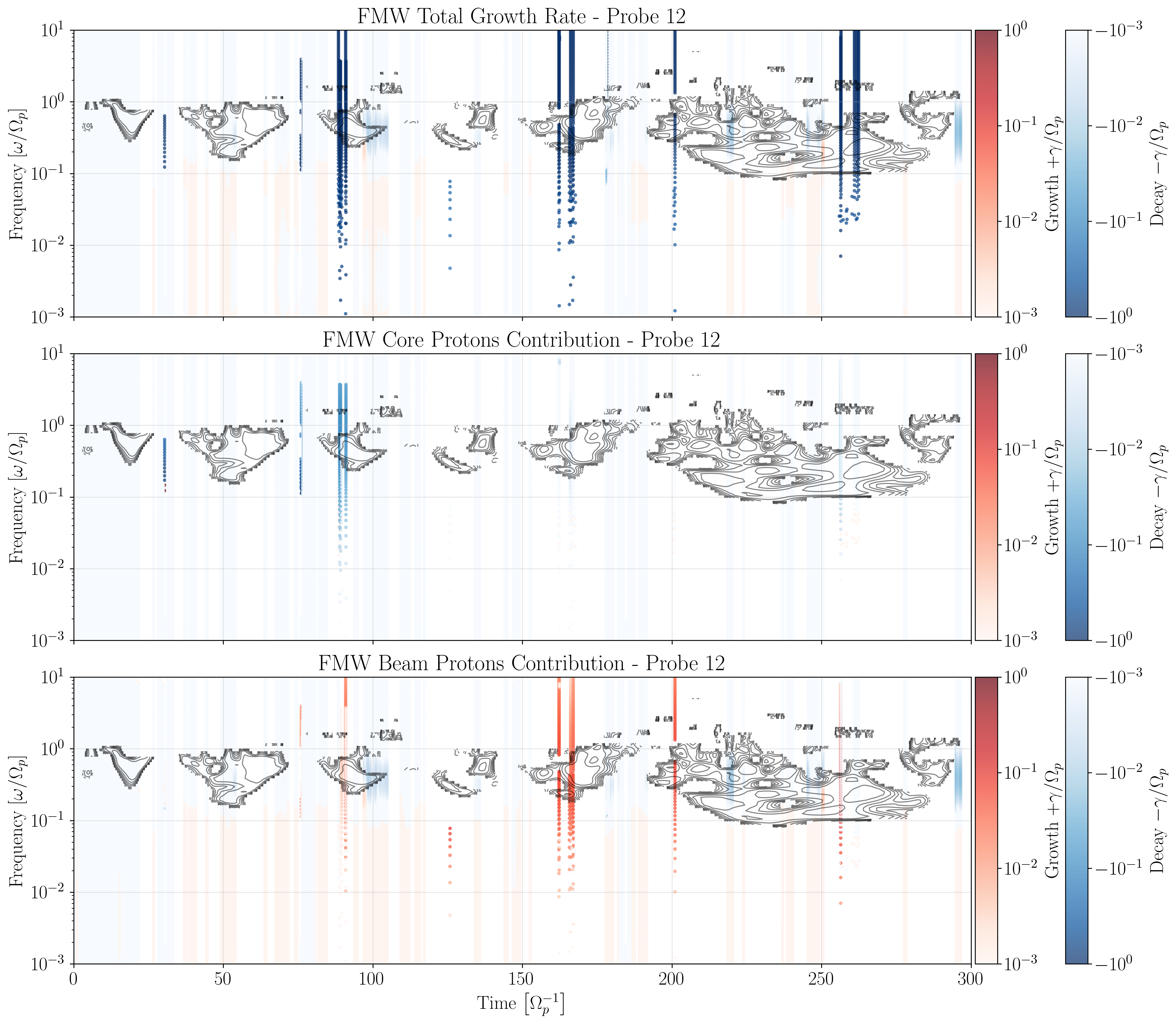}
    \caption{PLUME‑derived FMW growth rates decomposed by species. The total growth rate (top panel) is rarely positive, suggesting that the FMW are not driven by a linear instability but rather arise from nonlinear processes. The core contribution (second panel) is occasionally positive, but the total remains negative because the combined damping from the beam and electrons dominates. The beam contribution (third panel) is weaker, indicating that the beam does not drive the wave. The contours mark regions where the reduced magnetic helicity satisfies $\sigma_B<-0.7$, indicating the presence of right‑handed (FMW) wave activity.}
    \label{fig: FMW12gamma}
\end{figure}

For the FMW, the narrative is different. Figure~\ref{fig: FMW12gamma} presents the total FMW growth rate (top panel) from PLUME and the rates decomposed by species contribution. Unlike the ICW case, the PLUME solutions for the right‑handed FMW branch appear consistently weak across the spectrogram, with pale colors indicating that linear growth rates are very small or near zero at most times and frequencies. Although the waves are clearly present in the wavelet‑based helicity spectrogram, PLUME does not find a corresponding strong linear growth. This discrepancy suggests that the observed FMW are not well described by linear instability theory. 

Several nonlinear mechanisms from the literature can generate FMW without requiring local linear growth. In the monochromatic, large‑amplitude Alfvén wave simulations of \citet{2020ApJ...904...81G}, parametric decay produces daughter waves, including FMW, that are not linear eigenmodes of the initial pump. Phase steepening of broadband Alfvénic fluctuations, studied by \citet{2021ApJ...914L..36G}, leads to wave breaking and the formation of forced compressible perturbations of the fast type, characterized by right‑handed polarization and sharp density compressions in phase with magnetic pressure fluctuations \citep{2021ApJ...914L..36G, bianco2026evolution}.

A further indication that linear theory is insufficient is the breakdown of the species decomposition. In the PLUME output, the sum of the individual species contributions (\(\gamma_c + \gamma_b + \gamma_e\)) does not equal the total growth rate \(\gamma_{\text{tot}}\) at all times. By examining the ratio \(|\gamma/\omega_r|\) (where \(\omega_r\) is the real frequency), we find that the disagreement occurs when \(|\gamma/\omega_r| \gtrsim 1/e \approx 0.368\). In this regime, the mode is no longer weakly damped or growing, and the additive decomposition derived from the anti‑Hermitian part of the susceptibility tensor (which assumes \(|\gamma/\omega_r| \ll 1\)) becomes invalid \citep{2017JPlPh..83d5301K, 2021ApJ...909....7K}. In our simulation, the FMW growth/damping rates often exceed this threshold, indicating that the linear approximation is breaking down and that the waves are either strongly damped or nonlinearly generated. The combination of weak linear growth rates and the robust presence of FMW in the wavelet analysis indicates that these waves are likely generated nonlinearly and strongly damped, rather than arising from a linear instability. Consequently, we cannot reliably determine which species (core, beam, or electrons) drives the FMW solely from the linear PLUME analysis, and a simple linear description is insufficient for FMW in this simulation.

\begin{figure}
    \centering
    \includegraphics[width=1\linewidth]{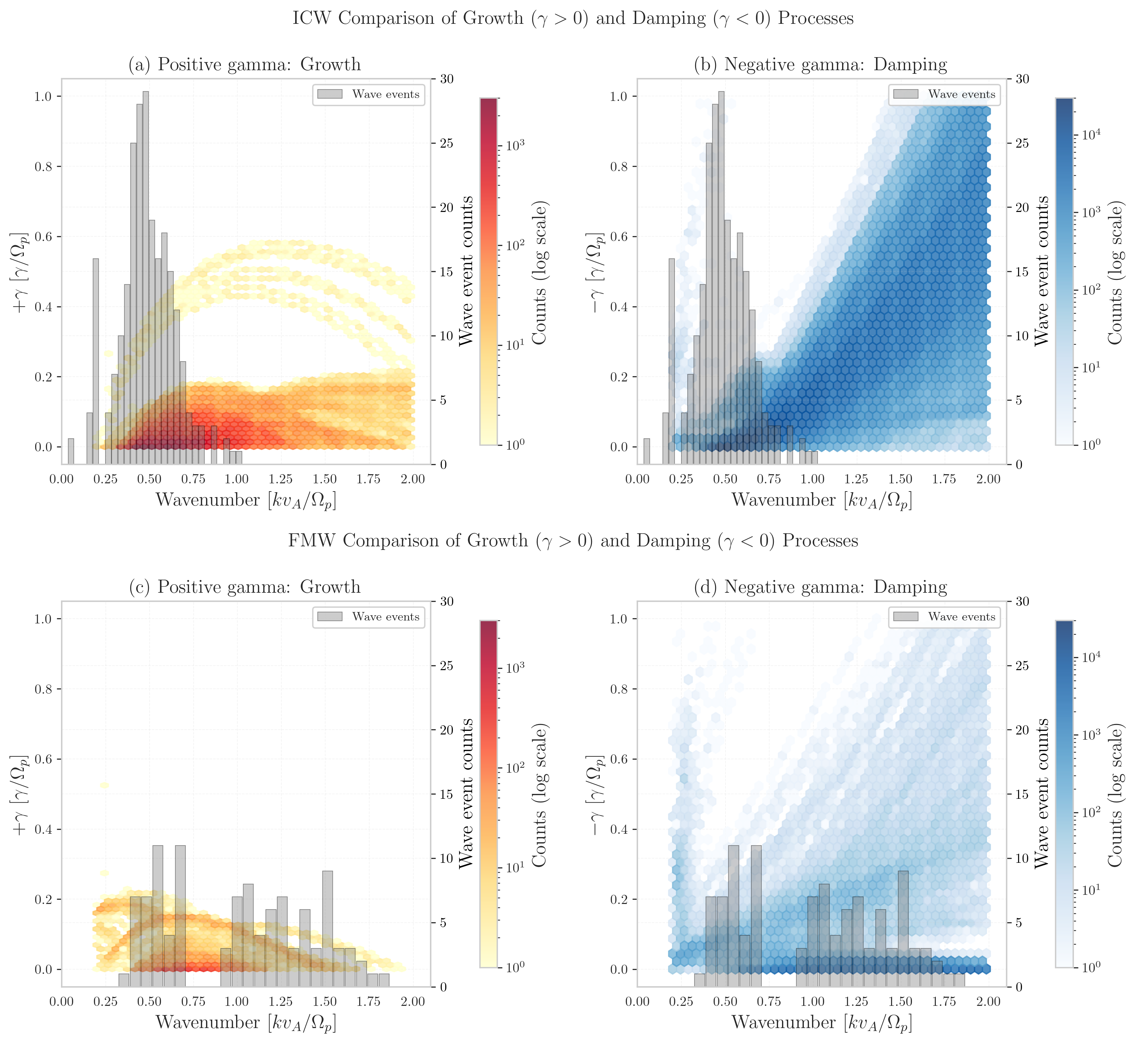}
    \caption{Distribution of growth rates ($10^{-3}<\frac{|\gamma|}{{\Omega_{p}}}<1$) in wavenumber space. The 2D histogram (color bar) shows the PLUME-derived growth rates, with positive values (red) indicating growth and negative values (blue) indicating damping. The black histogram shows the count of wave events identified from the reduced magnetic helicity spectrogram ($|\sigma_B| > 0.8$) binned in $k$-space. For ICW, the majority of wave events are concentrated at $\frac{kv_A}{\Omega_p} \approx 0.5$, while the densest positive gammas happen at $\frac{kv_A}{\Omega_p} \approx 0.65$ with values $\frac{\gamma}{\Omega_p} \approx 5\times10^{-2}$. For FMW, both the gamma values and wave event counts are significantly lower, with no clear correspondence between the two diagnostics. The location of the FMW wave event peak shows no concentration of positive gamma values, and damping dominates across the entire $k$-range.}
    \label{fig: gammak}
\end{figure}

Figure~\ref{fig: gammak} shows the gammas between $10^{-3}$ and $1$ ($10^{-3}<\frac{|\gamma|}{{\Omega_{p}}}<1$) on $k$ space. The data is displayed with a 2D histogram on the main axis. The color bars (log scale) indicate the number of PLUME solutions falling into each two‑dimensional bin ($\gamma$, $k$). We limit the displayed range of $|\gamma|/\Omega_p$ to be $<1$ because values with $|\gamma|/\Omega_p>1$ correspond to modes where the damping or growth rate exceeds the real frequency. Such modes are overdamped and do not propagate as coherent oscillations. The black histogram on the same plot shows the count of wave events identified from the reduced magnetic helicity spectrogram (Figure~\ref{fig: sigmaBE}), binned in $k$-space. The counts of the wave events are shown on the secondary vertical axis on the right.

For ICW, it is evident that the majority of the wave events are identified at around $\frac{kv_A}{\Omega_p}\approx 0.5$ (peak of the black histogram). The positive gammas mostly concentrated around $\frac{kv_A}{\Omega_p}\approx0.65$ with values at around $\frac{\gamma}{\Omega_p}\approx5\cdot10^{-2}$ (Figure~\ref{fig: gammak}a). The $k$ value where the most wave events happen does not coincide with the $k$ value for the densest gamma values because of strong damping at higher $k$. At higher $k$ values (around $\frac{kv_A}{\Omega_p}\approx0.65$ and above), damping becomes increasingly dominant (Figure~\ref{fig: gammak}b). Even though the linear growth rates are higher in these regions, the net effect includes strong damping that prevents waves from being observed. As a result, these wavenumbers are underrepresented in the helicity-based wave-event histogram. At lower $k$ values, damping is less pronounced, which explains why even modest positive gammas can still produce detectable wave events; in essence, we argue that the predominant population of observed waves occurs where the damping is the weakest.

FMW presents a distinctly different case. Both the gammas and the reduced magnetic helicity wave events show significantly lower counts compared to ICW, and there is no clear pattern matching the two diagnostics (Figure~\ref{fig: gammak}c). The location where the wave event histogram peaks corresponds to no noticeable concentration of positive gamma values. Across the entire $k$-range, the damping rates are generally high (Figure~\ref{fig: gammak}d). Yet despite this dominating linear damping, FMW are still detected in the reduced magnetic helicity spectrogram. This apparent contraction suggested that the observed FMW, for the most part, are not linearly generated modes but rather arise from nonlinear processes. The waves observed in the simulation may be nonlinear structures that persist despite linear damping, explaining why they appear in the helicity data while PLUME fails to identify corresponding linear modes.

The contrasting behavior of ICW and FMW in Figure~\ref{fig: gammak} highlights their fundamentally different roles in the plasma. ICW (Figure~\ref{fig: gammak}a,b) exhibit a clear correspondence between linear theory and simulation (the positive gammas align closely with the wavenumbers where wave events are detected), confirming that ICW are linear modes. The slight offset of $k$ in Figure~\ref{fig: gammak}a between the dense region of $\gamma$ and the peak of wave events reflects the influence of damping.

FMW, by contrast, shows no such correspondence (Figure~\ref{fig: gammak}c,d). Positive gamma peaks do not match the wave event peaks. This indicates that the FMW in the simulation is not a linear eigenmode but rather a nonlinear structure that persists despite linear damping. Since PLUME cannot reproduce the observed waves as linear eigenmodes, yet the waves are unambiguously present in the simulation, we attribute their origin to nonlinear generation mechanisms.

\section{Conclusion} \label{conclusion}
This study investigated the generation and evolution of ion-scale waves in a collisionless plasma using hybrid simulations that combined wavelet-based polarization analysis, bi-Maxwellian VDF fitting, and linear stability analysis with the PLUME solver. Our goal was to identify plasma waves in hybrid-kinetic simulations of large-amplitude Alfvénic fluctuations, which would help us understand the generation and role of wave-particle interactions in solar wind heating.

The analysis reveals two fundamentally distinct wave populations operating simultaneously. ICW are linear modes that exhibit reliable correspondence with theoretical dispersion relations. PLUME converges for all time steps, confirming that they are eigenmodes of the evolving plasma. The core proton population dominates ICW growth, with beam contribution remaining negligible. Intermittent ICW generation occurs when core temperature anisotropy peaks, likely resulting from anisotropy buildup at phase‑steepened Alfvénic fronts. Following each growth episode, the core anisotropy relaxes toward marginal stability. This behavior is consistent with the core‑driven, linear nature of ICW.

FMW present a contrasting picture. Although PLUME converges for all time steps, the $|\gamma_\text{tot}|/\Omega_p$ solutions for the FMW branch are consistently low, indicating very small or near-zero growth rates at most times and frequencies, and there is no correspondence between linear growth regions and observed wave events. The frequency mismatches between the PLUME solutions and the observed waves, together with the clear presence of FMW in the helicity spectrogram, suggest that these waves are not linear eigenmodes of the instantaneous plasma. Further evidence comes from the breakdown of the species decomposition: the sum of the individual species contributions ($\gamma_c + \gamma_b + \gamma_e$) does not equal the total growth rate $\gamma_{\text{tot}}$, indicating that the linear approximation is breaking down and the waves are either strongly damped or nonlinearly generated. However, the qualitative agreement with the cold-plasma dispersion $\omega(k)$, even in the presence of strong damping, is curious and warrants further investigation. A positive net energy flux ratio means the net energy flow remains forward at all scales. The combination of weak linear growth rates, the robust presence of FMW in the wavelet helicity spectrogram, and the breakdown of the linear species decomposition indicates that these waves are likely nonlinearly generated rather than arising from linear instability. Consequently, we cannot reliably determine which species drives the FMW solely from the linear PLUME analysis, and a simple linear description is insufficient to understand the growth and damping of the FMW in this simulation.

We note, however, that the bi-Maxwellian assumption may not fully capture the complexity of the velocity distributions, especially for the FMW case. A more sophisticated treatment of the VDFs (e.g., using the Arbitrary Linear Plasma Solver, ALPS) could provide further insight into the wave-particle interactions and possibly resolve some of the discrepancies between linear theory and the observed wave activity.

\begin{acknowledgements}
This research was supported by the International Space Science Institute (ISSI) in Bern, through ISSI International Team project \#612 (Excitation and Dissipation of Kinetic-Scale Fluctuations in Space Plasmas) led by K.~G.~Klein; D.V. is supported by STFC Consolidated Grant ST/W001004/1; CAG is supported by NSF SHINE grant \#80NSS\-C18K1211. We acknowledge the Texas Advanced Computing Center (TACC) at The University of Texas at Austin for providing HPC resources.  Simulations have been run on the Frontera supercomputer http://www.tacc.utexas.edu; This work was supported by NASA HSR grant 80NSSC24K0272 and the PSP mission contract NNN06AA01C.
\end{acknowledgements}

\bibliography{citations}{}
\bibliographystyle{aasjournalv7}



\end{document}